\documentclass[12pt,twoside]{article}

\usepackage[a4paper]{geometry}
\usepackage{lineno}

\makeatletter
\renewcommand\title[1]{\gdef\@title{\reset@font\Large\bfseries #1}}
\renewcommand\section{\@startsection {section}{1}{\z@}%
	{-3.4ex \@plus -1ex \@minus -.2ex}%
	{2.1ex \@plus.2ex}%
	{\normalfont\large\bfseries}}
\renewcommand\subsection{\@startsection{subsection}{2}{\z@}%
	{-2.8ex\@plus -1ex \@minus -.2ex}%
	{1.3ex \@plus .2ex}%
	{\normalfont\normalsize\bfseries}}
\def\@runningauthor{}\newcommand{\runningauthor}[1]{\def\runningauthor{#1}}
\def\@runningtitle{}\newcommand{\runningtitle}[1]{\def\runningtitle{#1}}
\renewcommand{\ps@plain}{%
	\renewcommand{\@evenfoot}{\footnotesize Sequences and Their Applications (SETA) 2026\hfill\thepage}
	\renewcommand{\@oddfoot}{\footnotesize Sequences and Their Applications (SETA) 2026\hfill\thepage}
	\renewcommand{\@evenhead}{\footnotesize\scshape \hfill\runningauthor\hfill}
	\renewcommand{\@oddhead}{\footnotesize\scshape \hfill\runningtitle\hfill}}
\g@addto@macro\bfseries{\boldmath}
\makeatother

\usepackage{amsthm,amsmath,amssymb,amsfonts,mathtools}
\usepackage{enumitem}
\usepackage{booktabs}
\usepackage{array}
\usepackage[colorlinks=true,citecolor=black,linkcolor=black,urlcolor=blue]{hyperref}
\usepackage[textsize=footnotesize]{todonotes}
\makeatletter
\define@key{todonotes}{magenta}[]{\setkeys{todonotes}{backgroundcolor=magenta!40}}
\makeatother

\theoremstyle{plain}
\newtheorem{theorem}{Theorem}[section]
\newtheorem{lemma}[theorem]{Lemma}
\newtheorem{proposition}[theorem]{Proposition}
\newtheorem{corollary}[theorem]{Corollary}

\newtheorem{definition}[theorem]{Definition}
\newtheorem{example}[theorem]{Example}
\newtheorem{remark}[theorem]{Remark}

\DeclareMathOperator{\E}{Ext}
\DeclareMathOperator{\Ext}{Ext}
\newcommand{\one}{\mathbf{1}}

\setlist{leftmargin=2.1em,itemsep=1pt,topsep=1pt,parsep=0pt}
\newcommand{\F}{\mathbb{F}}
\newcommand{\Ftwo}{\mathbb{F}_2}
\newcommand{\C}{\mathbb{C}}
\newcommand{\Wal}{W}
\newcommand{\rank}{\operatorname{rank}}

\newcommand{\Mb}{M_{\rm b}}

\newcommand{\ip}[2]{\langle #1,#2\rangle}

\title{Compatible permutations by Schur-complement extensions and their applications to NOMA codebooks and MUBs}
\runningtitle{Compatible permutations and Schur-complement extensions}
\runningauthor{C. Li, C. Riera, P. St\u{a}nic\u{a}}

\author{Chunlei Li\footnote{
		\small Department of Informatics, University of Bergen, Norway; 
		\small\tt chunlei.li@uib.no}
	\and
	Constanza Riera\footnote{
		\small Department of Computer Science, Electrical Engineering and Mathematical Sciences, 
		\small Western Norway University of Applied Sciences, Bergen, Norway; 
		\small\tt csr@hvl.no}
	\and
	Pantelimon St\u{a}nic\u{a}\footnote{
		\small Applied Mathematics Department, Naval Postgraduate School, Monterey, CA, USA; 
		\small\tt pstanica@nps.edu}}
\date{}

\begin{document}
	\maketitle
	\thispagestyle{empty}

	\begin{abstract}
		For a permutation $\pi$ of $\{1,\dots,n\}$, let $Q_\pi(x)=\sum_{i=1}^{n-1}x_{\pi(i)}x_{\pi(i+1)}$ be the quadratic Boolean function of the Hamiltonian path $\pi(1)-\pi(2)-\cdots-\pi(n)$. Two permutations $\pi,\sigma$ are \emph{compatible} if $Q_\pi+Q_\sigma$ is bent (for even $n$) or near-bent (for odd $n$). Large sets of pairwise compatible permutations give binary codebooks with low peak-to-average power ratio (PAPR) and low coherence for grant-free non-orthogonal multiple access (NOMA). In even dimension, a compatible set of size $L$ also gives $L$ mutually unbiased bases.
		
		We construct compatible sets of size $18$ in every even dimension $n\ge10$ except $n=14$, and sets of size $15$ in dimensions $8$ and $14$. The construction uses a Schur-complement criterion for repeated right extensions. Three inverse endpoint entries for each pair of paths determine whether compatibility persists at every iteration. An endpoint lift gives the same lower bounds in the next odd dimensions, and a cyclic construction gives $28$ compatible permutations in dimension $7$. In dimension six the maximum is nine, or $8$ under the requirement that every self-extension be compatible. In odd dimension $n\ge5$ we also prove the size upper bound $\binom n2(n-2)$.
	\end{abstract}
	
	\maketitle
	
	\section{Introduction}
	Let $S_n$ be the set of all permutations on $\{1,2,\dots, n\}$ for an integer $n\geq 2$.
	A permutation $\pi=[\pi(1),\ldots,\pi(n)]\in S_n$ determines a Hamiltonian path $\Gamma_\pi=\pi(1)-\cdots-\pi(n)$ and the quadratic Boolean function $Q_\pi(x)=\sum_{i=1}^{n-1}x_{\pi(i)}x_{\pi(i+1)}$ on $\Ftwo^n$.  Two permutations $\pi, \sigma \in S_n$ are said to be \textit{compatible} when the function $Q_\pi+Q_\sigma$ is bent for even $n$ and when the sum is near-bent for odd $n$.  
	A set of mutually compatible permutations is termed a \textit{compatible set}.
	A compatible set of size $L$ gives a codebook with $L2^n$ spreading sequences for grant-free code-domain non-orthogonal multiple access (NOMA)~\cite{PatersonSETA2002,yulnomacodebook,Liu2024,Tian,CCDS25}. In even dimension, the sequences form $L$ mutually unbiased bases~\cite{MUBs}. Pairwise bent sums also occur in constructions from Kerdock codes and symplectic spreads~\cite{Calderbank1997}. Here the restriction to quadratic functions supported on Hamiltonian paths ensures that each spreading sequence has peak-to-average power ratio (PAPR) at most two~\cite{DavisJedwab1999}.
	
	Paterson also studied the low-PAPR condition for multicode CDMA~\cite{Paterson2004}. Let $M(n)$ denote the largest size of a compatible set in $S_n$. Paterson~\cite{PatersonSETA2002} proved that $M(n)\le\binom n2$ for even $n$, with equality at $n=4$.
	We seek large compatible sets in every dimension $n\ge6$. Section~\ref{sec:2} relates compatibility to the rank of a binary adjacency matrix and records the resulting codebook parameters. Section~\ref{sec:3} uses a Schur complement to test compatibility after adding vertices to two paths. For repeated right extensions by fixed blocks, Theorem~\ref{Th_mix_extension} reduces the infinite sequence of tests to two products of inverse endpoint entries for each pair.
	
	Section~\ref{sec:constructions} uses this criterion to construct compatible sets of size $18$ in every even dimension $n\ge10$ except $n=14$, and sets of size $15$ in dimensions $8$ and $14$. The seeds and blocks are arranged in cyclic orbits to reduce the number of comparisons in the search. We also establish the exact maxima in dimension six for ordinary compatibility and for self-extension. Section~\ref{sec:odd} transfers the even-dimensional lower bounds to the next odd dimension, gives a separate construction of size $28$ in dimension $7$, and proves an upper bound for odd dimensions. Table~\ref{tab:bounds} summarizes the bounds.
	
	\begin{table}[!t]
		\centering\small
		\caption{Updated lower bounds on the maximum size $M(n)$}
		\label{tab:bounds}
		\begin{tabular}{@{}llp{.34\linewidth}@{}}
			\toprule
			Dimension $n$ & Lower bound & Reference\\
			\midrule
			% $n=6$ & $9$ & Proposition~\ref{prop:sixexact}\\
			$n=7$ & $28$ & Corollary~\ref{cor:uniform15}\\
			$n\in\{8,9,14,15\}$ & $15$ & Corollary~\ref{cor:uniform15}\\
			$n\ge10$, $n\notin\{14,15\}$ & $18$ & Corollary~\ref{cor:uniform15}\\
			$n\equiv0\pmod4$ & $6$ & \cite{CCDS25}\\
			$n$ even & $(p-1)/2$, $p\mkern2mu\mid\mkern2mu n+1$ least prime & \cite{Liu2024}\\
			$n$ odd & $(p-1)/2$, $p\mid n$ least prime & \cite{Liu2024}\\
			\bottomrule
		\end{tabular}
	\end{table}
	
	\section{Preliminaries}\label{sec:2}
	
	\subsection{Notation and basic results}
	
	We write permutations as sequences, e.g., $\pi=[\pi(1),\dots,\pi(n)]\in S_n$. We write $[n]=\{1,\ldots,n\}$, denote the identity permutation by $I_n=[1,2,\ldots,n]$, and write $\tilde\pi=[\pi(n),\ldots,\pi(1)]$ for the reverse of $\pi$. Composition is $(\sigma\circ\pi)(i)=\sigma(\pi(i))$. 
	
	\begin{definition}\label{def:Qpi}
		For a permutation $\pi\in S_n$, let $\Gamma_\pi$ be a graph  of vertex set $\{1,2,\ldots,n\}$ and edge set $E(\Gamma_\pi)=\bigl\{\{\pi(i),\pi(i+1)\}:1\le i\le n-1\bigr\}$.
		The graph $\Gamma_\pi$ is a Hamiltonian path of the form ${\pi(1)} 
		- \pi(2) - \cdots -  {\pi(n)}$.%, simply as $(\pi(1), \pi(2), \cdots, \pi(n))$. 
	\end{definition}
	The sequences $\pi$ and $\tilde\pi$ are the two orientations of the same undirected path. They determine the same quadratic function, although their orientations matter when we specify an endpoint for an extension.
	\begin{definition}\label{def:extension}
		Let $\pi\in S_n$, $\rho\in S_m$ and $0\le r\le m$. The \emph{extension of $\pi$ by $\rho$ at position $r$} is
		\[
		\Ext^r_\rho(\pi)=[\,\rho(1)+n,\dots,\rho(r)+n,\ \pi(1),\dots,\pi(n),\ \rho(r+1)+n,\dots,\rho(m)+n\,]\in S_{n+m}.
		\] In particular,  $	\Ext^0_\rho(\pi) = [\pi(1),\dots,\pi(n),\ \rho(1)+n,\dots,\rho(m)+n]$.
	\end{definition}
	
	For an integer $j$ we write $\rho+j=[\rho(1)+j,\dots,\rho(m)+j]$ as a $j$-translate of $\rho\in S_m$.
	Denote by $\|$ the concatenation of sequences. 
	Thus $\pi$ occupies positions $r+1,\dots,r+n$ of $\Ext^r_\rho(\pi)$, the new vertices $n+1,\dots,n+m$ appear in the order prescribed by $\rho$, and $\Ext^0_\rho(\pi)=\pi\|(\rho+n)$. For instance, $\Ext^1_{[1,2]}([2,1,3,4])=[5,2,1,3,4,6]$. 
	For $\rho\in S_t$ and $k\ge1$, the \emph{self-extension} of $\rho$ is
	\[
	\rho^{(k)}=\rho\,\|\,(\rho+t)\,\|\,\cdots\,\|\,(\rho+(k-1)t)\in S_{tk},
	\]
	and $\rho^{(0)}$ is the empty sequence. For example, $[1,3,4,2]^{(2)}=[1,3,4,2,5,7,8,6]$.
	
	All adjacency matrices and their block matrices are over $\Ftwo$. We denote by $e_i$ the $i$-th standard column vector (its size will be clear from the context), by $\one$ the all-one vector, by $E_{i,j}=e_ie_j^\intercal$ the matrix unit, and by $I_t$ the $t\times t$ identity matrix. 
	
	The \emph{adjacency matrix} $A_\pi$ of the Hamiltonian path $\Gamma_{\pi}$ for $\pi\in S_n$ is the $n\times n$ matrix with $(A_\pi)_{u,v}=1$ if and only if $\{u,v\}$ is an edge of $\Gamma_\pi$. It is symmetric with zero diagonal, and $A_\pi=A_{\tilde\pi}$. A square matrix $M$ over $\Ftwo$ is \emph{alternating} if it is symmetric with zero diagonal. Equivalently, $x^\intercal Mx=0$ for all $x$.
	
	We use the following Schur-complement identities~\cite{Zhang2005} to test the rank of extended adjacency sums.  
\begin{lemma}\label{lem_SC}
Let $M=\begin{bmatrix}A&B\\C&D\end{bmatrix}$ be a square block matrix over $\F_2$ with $A$ nonsingular (hence $\det(A)=1$), and let $S=D+CA^{-1}B$ be the Schur complement of $A$ in $M$. Then $\det(M)=\det(S)$. If $S$ is nonsingular, then
\begin{equation}\label{eq:block-inverse}
M^{-1}=\begin{bmatrix}A^{-1}+A^{-1}BS^{-1}CA^{-1} & A^{-1}BS^{-1}\\ S^{-1}CA^{-1} & S^{-1}\end{bmatrix}.
\end{equation}
\end{lemma}
\begin{proof}
Both claims follow from the factorization
\[
M=\begin{bmatrix}I&0\\CA^{-1}&I\end{bmatrix}\begin{bmatrix}A&0\\0&S\end{bmatrix}\begin{bmatrix}I&A^{-1}B\\0&I\end{bmatrix}.
\]
Both outer factors are block triangular with identity diagonal blocks and hence have determinant one. Since $A$ is nonsingular over $\Ftwo$, its determinant is also one. The factorization therefore gives $\det(M)=\det(S)$. If $S$ is nonsingular, the middle factor has inverse $\operatorname{diag}(A^{-1},S^{-1})$. Each outer factor is inverted by negating its off-diagonal block, which has no effect in characteristic two. Multiplying the three inverses in reverse order gives~\eqref{eq:block-inverse}.
	\end{proof} 
	
	We call $CA^{-1}B$ the \emph{offset matrix} in the Schur complement $S=D+CA^{-1}B$.

	\subsection{Quadratic forms and compatibility}
	
	The ranks of the symplectic matrices determine the Walsh spectra of quadratic Boolean functions. For path-graph functions these matrices are the path adjacency matrices.
	
	For an $n$-variable Boolean function $f:\Ftwo^n\to\Ftwo$, its Walsh-Hadamard   transform at $c
	\in \F_2^n$ is given by $\Wal_f(c)=\sum_{x\in\Ftwo^n}(-1)^{f(x)+c\cdot x}$.  The function $f$ is bent when $n$ is even and $|\Wal_f(c)|=2^{n/2}$ for all $c\in \F_2^n$, and near-bent when $n$ is odd and $\Wal_f(c)\in\{0,\pm2^{(n+1)/2}\}$ for all $c\in \F_2^n$.  
	For a quadratic Boolean function $Q$, let $C_Q$ be the strictly upper triangular $n\times n$ matrix whose $(i,j)$ entry for $i<j$ is the coefficient of $x_ix_j$ in the algebraic normal form of $Q$. Its symplectic matrix is $B_Q=C_Q+C_Q^\intercal$, where $\intercal$ denotes transpose. The Walsh-Hadamard  spectrum of quadratic Boolean functions can be characterized in terms of their symplectic matrices as in the following lemma (see, for example,~\textup{\cite[p.~441]{MacWilliams1977}}).
	\begin{lemma}\label{prop:rank}
		Let $Q: \Ftwo^n \rightarrow \Ftwo$ be a quadratic Boolean function with $\rank(B_Q)=r$. Then the Walsh-Hadamard  spectrum $\{\Wal_Q(c):c\in \F_2^n\} \subseteq \{0,\pm2^{(2n-r)/2}\}$.  In particular, $Q$ is bent exactly when $n$ is even and $r=n$, and near-bent exactly when $n$ is odd and $r=n-1$. 
	\end{lemma}
	
	\begin{proof}
		Let $R=\ker B_Q$. For $x,u\in\Ftwo^n$, polarization gives $Q(x+u)+Q(x)=Q(u)+Q(0)+x^\intercal B_Q u$. Squaring the Walsh sum and substituting $x+u$ for the second summation variable therefore gives
		\[
		\Wal_Q(c)^2
		=2^n\sum_{u\in R}(-1)^{Q(u)+Q(0)+c\cdot u}.
		\]
		Indeed, summing $(-1)^{x^\intercal B_Qu}$ over $x$ gives $2^n$ on $R$ and zero outside $R$. The function $u\mapsto Q(u)+Q(0)$ is linear on $R$ because its polar form vanishes there. Thus the last sum is $|R|=2^{n-r}$ if this linear function agrees with $u\mapsto c\cdot u$ on $R$, and zero otherwise. This proves the asserted spectrum. A linear functional on $R$ extends to the ambient space, so at least one Walsh value is nonzero. When $r=n$, the radical is zero and every Walsh value has absolute value $2^{n/2}$. When $n$ is odd and $r=n-1$, the nonzero values have absolute value $2^{(n+1)/2}$. Conversely, each prescribed nonzero amplitude forces the corresponding value of $r$.
	\end{proof}
	
	We call the quadratic function $Q_\pi(x)=\sum_{i=1}^{n-1}x_{\pi(i)}x_{\pi(i+1)}$ the 
	\emph{path-graph function} of a permutation $\pi$.
	Reversing the path preserves $Q_\pi$. Each edge contributes one quadratic monomial, so $B_{Q_\pi}=A_\pi$. For a pair of paths this gives $B_{Q_\pi+Q_\sigma}=A_\pi+A_\sigma$. The Walsh criterion therefore gives the following equivalent formulation of compatibility.
	
	\begin{definition}\label{def:compatible}
		Two permutations $\pi,\sigma\in S_n$ are \textit{compatible}, written
		$\pi\sim\sigma$, if $Q_\pi+Q_\sigma$ is bent when $n$ is even, or near-bent
		when $n$ is odd. Alternatively, $\pi\sim\sigma$ 
		if and only if the matrix sum $A_\pi+A_\sigma$ is nonsingular for even $n$, or $A_\pi+A_\sigma$ has rank $n-1$ for odd $n$.  
	\end{definition}
	
	\begin{lemma}\label{lem:symmetries}\textup{\cite{CCDS25}}
		For any $\pi,\sigma\in S_n$, the following equivalences hold.
		\begin{enumerate}[label=$(\roman*)$]
			\item $\sigma\sim\pi$ iff $\sigma\sim\tilde\pi$ \quad (reversal symmetry).
			\item $I_n\sim\pi$ iff $I_n\sim\pi^{-1}$  \quad (inverse symmetry).
			\item $\pi\sim\sigma$ iff $I_n\sim\sigma^{-1}\circ\pi$ \quad (composition).
		\end{enumerate}
	\end{lemma}
	\begin{proof}
		Path reversal leaves $A_\pi$ unchanged, which proves (i). For a relabeling $\lambda\in S_n$, simultaneous relabeling of both paths conjugates their adjacency sum by the permutation matrix of $\lambda$. It therefore preserves its rank. Taking $\lambda=\sigma^{-1}$ proves (iii). Relabeling the pair $(I_n,\pi)$ by $\pi^{-1}$ gives $(\pi^{-1},I_n)$, which has the same compatibility as $(I_n,\pi^{-1})$ because adjacency matrix addition is commutative. This proves (ii).
	\end{proof}
	
	From Lemma~\textup{\ref{lem:symmetries}}, it follows that a set $S\subset S_n$ is compatible iff so is
	$S' = \{\pi^{-1}\circ\sigma : \sigma\in S\}$ for any fixed $\pi\in S$.
	Hence	
	we may normalize any compatible family to contain $I_n$. Its other members form a pairwise compatible subset of $IS_n=\{\pi\in S_n:\pi\sim I_n\}$. Conversely, adjoining $I_n$ to any such subset gives a compatible family.
	
	\subsection{Codebooks and mutually unbiased bases}
	
	The Walsh spectrum of each pairwise sum determines the inner products between the spreading sequences from the two paths. For $\pi\in S_n$ and $a\in\Ftwo^n$, let $s_{\pi,a}=2^{-n/2}\bigl((-1)^{Q_\pi(x)+a\cdot x}\bigr)_{x\in\Ftwo^n}\in\C^{2^n}$, with the coordinates ordered lexicographically. For a set $S\subseteq S_n$, let $\Phi_S$ be the $2^n\times |S|2^n$ matrix with columns $s_{\pi,a}$, $\pi\in S$, $a\in\Ftwo^n$, and let $\mu(\Phi_S)$ be the largest absolute inner product of two distinct columns~\cite{yulnomacodebook,CCDS25}.
	
	For a column $s=(s_0,\ldots,s_{N-1})$, the PAPR is the maximum of $\bigl|\sum_{j=0}^{N-1}s_jz^j\bigr|^2$ over $|z|=1$, divided by its mean power $\sum_{j=0}^{N-1}|s_j|^2$. Two orthonormal bases in dimension $N$ are mutually unbiased if every inner product between vectors from the two bases has absolute value $N^{-1/2}$.
	
	\begin{proposition}\label{prop:codebook}
		Let $S\subseteq S_n$ be a compatible set of size $L\ge2$. Then $\Phi_S$ is a $2^n\times L2^n$ binary codebook whose columns have PAPR at most $2$, and
		\[
		\mu(\Phi_S)=\begin{cases}2^{-n/2}, & n \text{ even},\\ 2^{-(n-1)/2}, & n\text{ odd}.\end{cases}
		\]
		For even $n$, the $L$ bases $\{s_{\pi,a}:a\in\Ftwo^n\}$, $\pi\in S$, of $\C^{2^n}$ are pairwise mutually unbiased.
	\end{proposition}
	
	\begin{proof}
		The binary phase sequence of each column is one member of a Golay pair~\cite{DavisJedwab1999,yulnomacodebook,CCDS25}. After applying the same normalization to the complementary sequence, the sum of the two squared polynomial magnitudes is two on the unit circle. This identity follows because the nonzero-lag autocorrelations of the two sequences cancel and each sequence has mean power one. The polynomial magnitude of the column therefore has square at most two, which gives the PAPR bound.
		For a fixed $\pi$, the columns $s_{\pi,a}$, $a\in\Ftwo^n$, form an orthonormal basis, since $\ip{s_{\pi,a}}{s_{\pi,b}}=2^{-n}\sum_x(-1)^{(a+b)\cdot x}$. For $\pi\ne\sigma$ in $S$,
		\[
		|\ip{s_{\pi,a}}{s_{\sigma,b}}|=2^{-n}\,|\Wal_{Q_\pi+Q_\sigma}(a+b)|.
		\]
		For even $n$, compatibility makes the right-hand side $2^{-n/2}$ for every $a,b$, which gives both the coherence and mutual unbiasedness. For odd $n$, the right-hand side is at most $2^{-(n-1)/2}$. The Walsh spectrum has a nonzero value by Lemma~\ref{prop:rank}, and as $a,b$ range over $\Ftwo^n$, their sum attains its index. This gives equality in the coherence bound.
	\end{proof}
	
	\begin{remark}
		\label{rem:graph-states}
		In quantum information theory, the path graph $\Gamma_\pi$ defines an
		$n$-qubit graph state $|G_\pi\rangle$. The quadratic phase
		$(-1)^{Q_\pi(x)}$ is the phase polynomial associated with that graph state in
		the computational basis. 
		For even $n$, the coherence $2^{-n/2}$ is the smallest possible for any two distinct orthonormal bases of $\C^{2^n}$. Earlier work on mutually unbiased measurements includes~\cite{Ivanovic1981}. At most $2^n+1$ pairwise mutually unbiased bases exist in $\C^{2^n}$, and complete sets of this size are known~\cite{Bandyopadhyay2002,KlappeneckerRoetteler2004,WoottersFields}. For the path forms considered here, Paterson's bound gives at most $\binom n2$ pairwise mutually unbiased phase bases. The restriction comes from the Hamiltonian paths used to obtain the Golay sequences.
	\end{remark}
	
	\section{
		A Schur-complement criterion for compatible extended permutations }\label{sec:3}
	
	The Schur complement isolates the effect of adding vertices to the paths. We first give the block form of the adjacency matrix for an extension as in Definition~\ref{def:extension}.
	
	\begin{lemma}\label{lem_adjmat}
		Let $\pi\in S_n$, $\rho\in S_m$, and $0\le r\le m$. Put $\vartheta=\Ext^r_\rho(\pi)$. Its adjacency matrix has the block form 
		\[
		A_{\vartheta}  = \begin{bmatrix}
			A_{\pi} & V \\
			V^\intercal & W
		\end{bmatrix}	
		\] where the matrices $V\in \F_2^{n\times m}$ and $W\in \F_2^{m\times m}$ are given by
		\[
		\begin{array}{|c|c|c|}
			\hline 
			& V & W \\ \hline 
			r = 0	 &E_{\pi(n), \rho(r+1)} & A_{\rho}  \\ 
			r = m	  & E_{\pi(1), \rho(r)} & A_{\rho}  \\ 
			0<r < m	&E_{\pi(n), \rho(r+1)} + E_{\pi(1), \rho(r) } & A_{\rho}  + E'_{\rho(r),\rho(r+1)} + E'_{\rho(r+1),\rho(r) }  \\ 
			\hline  
		\end{array}
		\]
		where $E_{i,j}\in \F_2^{n\times m}$ and $E'_{j,k}\in \F_2^{m\times m}$ are matrix units, taking one only at the entries $(i,j)$, $(j,k)$, respectively.
	\end{lemma}
	
	\begin{proof}
		The nonzero entries of $A_\vartheta$ occur in symmetric pairs
		$A_\vartheta(\vartheta(i),\vartheta(i+1))=A_\vartheta(\vartheta(i+1),\vartheta(i))=1$
		for $i=1,2,\ldots,n+m-1$.
		By the relation $(\vartheta(r+1),\dots,\vartheta(r+n)) = (\pi(1), \dots, \pi(n))$ for a certain integer $r$, 
		we have that for $i=1,\dots, n-1$,
		\[
		A_{\vartheta}(\vartheta(r+i),\vartheta(r+i+1))
		=A_{\vartheta}(\pi(i),\pi(i+1))=1,
		\quad
		A_{\vartheta}(\pi(i+1),\pi(i))=1.
		\]
		Since $\pi$	is a permutation on $\{1,\dots, n\}$, the top-left $n\times n$ sub-matrix of $A_{\vartheta}$ is actually the adjacency matrix $A_{\pi}$ which has $A_{\pi}(\pi(i),\pi(i+1))=A_{\pi}(\pi(i+1),\pi(i))=1$ for $i=1,\dots,n-1$.
		The position $r$ determines the edges between the old and new vertices and the edges within the new vertex set.
		
		When $r=0$, the added vertices occur consecutively on the right in the order prescribed by $\rho$, so $W=A_\rho$. The only edge between the two vertex sets joins $\pi(n)$ to $\rho(1)+n$, giving $V=E_{\pi(n),\rho(1)}$.
		
		Similarly, when $r=m$, we have $W= A_{\rho}$ and that the matrix $V$ satisfies $V(i,j)=1$ when $(i,j) = (\pi(1), \rho(m))$ and $V(i,j)=0$ otherwise, implying $V=E_{\pi(1),\rho(m)} $.
		
		When $0 < r < m$, we need to consider two pairs $(\vartheta(r), \vartheta(r+1)) = (\rho(r)+n, \pi(1)) $ and $(\vartheta(r+n), \vartheta(r+n+1)) = (\pi(n), \rho(r+1)+n) $ for the matrix $V$.
		In this case the matrix $V$ is the sum of the matrix units $E_{\pi(1), \rho(r)}$ and $E_{\pi(n), \rho(r+1)}$.
		Accordingly, the matrix $W$ is the matrix from $A_{\rho}$ in $\F_2^{m\times m}$ by flipping one at the entries $(\rho(r), \rho(r+1))$ and $(\rho(r+1), \rho(r))$ to zero, 
		i.e., $W = A_{\rho}  + E'_{\rho(r), \rho(r+1)} + E'_{\rho(r+1), \rho(r)} $.
	\end{proof}
	
	Lemma~\textup{\ref{lem_adjmat}} shows the block structure of  the adjacency matrix of an extension permutation $\vartheta=\Ext^r_\rho(\pi)$.
	Based on this block structure, we derive a criterion 
	on the compatibility of two extension permutations via Schur complements as below.
	\begin{lemma}\label{thm_comp_extension_mat}
		For even $n,m$, two compatible permutations $\pi_1, \pi_2 \in S_n$ and two permutations $\rho_1, \rho_2 \in S_m$,
		assume the two extension permutations $\vartheta_i=\Ext^{r_i}_{\rho_i}(\pi_i)$ at position $r_i$ for $i=1,2$  have adjacency matrices $A_{\vartheta_1}$, $A_{\vartheta_2}$, 
		(as given in Lemma~\textup{\ref{lem_adjmat}}), respectively.
		Then the permutations $\vartheta_1$ and $\vartheta_2$	are compatible if and only if 
		the Schur complement of $A_{\pi_1}+ A_{\pi_2}$ in the matrix $A_{\vartheta_1} + A_{\vartheta_2}$, given by
		$$(V_1+V_2)^\intercal (A_{\pi_1}+ A_{\pi_2})^{-1}(V_1+V_2)+ (W_1+W_2),$$ 
		is nonsingular.
	\end{lemma}
	
	\begin{proof}
		By Lemma~\textup{\ref{lem_adjmat}},
		\[
		A_{\vartheta_1}+A_{\vartheta_2}
		=
		\begin{bmatrix}
			A_{\pi_1}+A_{\pi_2} & V_1+V_2\\
			(V_1+V_2)^\intercal & W_1+W_2
		\end{bmatrix}.
		\]
		Since $\pi_1$ and $\pi_2$ are compatible, the matrix
		$A_{\pi_1}+A_{\pi_2}$ is nonsingular.  The Schur-complement criterion therefore indicates that
		$A_{\vartheta_1}+A_{\vartheta_2}$ over $\F_2$ is nonsingular if and only if
		$(W_1+W_2)+(V_1+V_2)^\intercal
		(A_{\pi_1}+A_{\pi_2})^{-1}(V_1+V_2)$
		is nonsingular. 
		By Lemma~\ref{prop:rank}, nonsingularity of the adjacency sum is equivalent to compatibility of $\vartheta_1$ and $\vartheta_2$.
	\end{proof}
	
	\begin{remark}
		Li et al.~\cite{CCDS25} obtained a Walsh-transform criterion for a right extension to be compatible with the identity permutation. Lemma~\ref{thm_comp_extension_mat} allows arbitrary insertion positions and compares the two extended paths. 
	\end{remark}
	
	The following example shows that a Schur complement can remain unchanged for properly chosen extending permutations.
	
	\begin{example}\label{ex:four-block-repeat}
		Let
		\(
		\pi_1=[3,2,4,1]\) and \(
		\pi_2=[3,4,1,2].
		\)
		Consider the extensions obtained by appending one translated copy of the same
		pattern to the right, namely
		\[
		\pi_1^{(2)}=[3,2,4,1,7,6,8,5],
		\qquad
		\pi_2^{(2)}=[3,4,1,2,7,8,5,6].
		\]
		These are self-extended permutations with $n=4$.  In the sum
		\[
		A_{\pi_1^{(2)}}+A_{\pi_2^{(2)}}=
		\begin{bmatrix}
			A_{\pi_1}+A_{\pi_2}&V_1+V_2\\
			(V_1+V_2)^{\intercal}&A_{\pi_1}+A_{\pi_2}
		\end{bmatrix},
		\]
		one has $V_1+V_2=E_{1,3}+E_{2,3}$. Moreover, taking $A=A_{\pi_1}+A_{\pi_2}$, we have
		\[
		A=
		\begin{bmatrix}
			0&1&0&0\\
			1&0&1&1\\
			0&1&0&1\\
			0&1&1&0
		\end{bmatrix}
		\text{ and }
		A^{-1}=
		\begin{bmatrix}
			0&1&1&1\\
			1&0&0&0\\
			1&0&0&1\\
			1&0&1&0
		\end{bmatrix}.
		\]
		A direct multiplication gives
		\[
		(E_{1,3}+E_{2,3})^{\intercal}A^{-1}(E_{1,3}+E_{2,3})=\mathbf{0}_{4\times4}.
		\] Hence the Schur complement of $A$ in $A_{\pi_1^{(2)}}+A_{\pi_2^{(2)}}$ is given by
		\begin{equation}\label{Eq_SC_unchanged}
			A + (E_{1,3}+E_{2,3})^{\intercal}A^{-1}(E_{1,3}+E_{2,3}) = A, 
		\end{equation}
		implying that the two extended
		permutations $\pi_1^{(2)}$ and $\pi_2^{(2)}$ are compatible.  Moreover, the same calculation repeats at every
		subsequent block. For self-extended permutations $\pi_i^{(k)}$, the lower-right $4\times 4$ block in the inverse after the $k$th stage remains
		$A^{-1}$ by~\eqref{eq:block-inverse}, so the next Schur complement is again $A$.  This property implies that
		$\pi_1^{(k)}$ and $\pi_2^{(k)}$ are compatible for all $k\ge1$.
	\end{example}
	
	In the sequel we will consider a more general extension, covering self-extensions as a special case.
	
	% \subsection{Iterative right extensions}\label{sec:right}
	
	For the rest of this section, we consider repeated right extensions in even dimensions. The next lemma shows how the two connecting edges enter the Schur complement.
	
	\begin{lemma}
		\label{lem:offset}
		Let $n,t$ be two positive integers and $M\in\F_2^{n\times n}$ be a nonsingular symmetric matrix with zero diagonal. Then $M^{-1}$ is also symmetric with zero diagonal. Moreover, for any column vectors $x_1,x_2\in\F_2^{n}$ and $y_1,y_2\in\F_2^{t}$ and the matrix $Z=x_1y_1^{\intercal}+x_2y_2^{\intercal}\in\F_2^{n\times t}$, one has
		\begin{equation}\label{Eq_offset}
			Z^{\intercal}M^{-1}Z=(x_1^{\intercal}M^{-1}x_2)\,\bigl(y_1y_2^{\intercal}+y_2y_1^{\intercal}\bigr).
		\end{equation}
	\end{lemma}
	
	\begin{proof}
		Symmetry follows from $(M^{-1})^{\intercal}=(M^{\intercal})^{-1}=M^{-1}$. Note that for any $x\in\F_2^n$ one has $x^{\intercal}Mx=\sum_{i}M_{i,i}x_i^2+2\sum_{i<j}M_{i,j}x_ix_j=0.$ 
		Letting $e_i$ be the $i$-th standard vector in $\F_2^n$, we have 
		$$
		(M^{-1})_{i,i}=e_i^{\intercal}M^{-1}e_i=
		(M^{-1}e_i)^{\intercal} M (M^{-1}e_i) = 0
		$$ for $i=1,\dots,n$, implying that $M^{-1}$ has zero diagonal.
		Furthermore, expanding  $Z^{\intercal}M^{-1}Z$ yields
		\[
		Z^{\intercal}M^{-1}Z=\sum_{i,j\in\{1,2\}}(x_i^{\intercal}M^{-1}x_j)\,y_iy_j^{\intercal} = (x_1^{\intercal}M^{-1}x_2)\,\bigl(y_1y_2^{\intercal}+y_2y_1^{\intercal}\bigr).
		\]
		since $x_1^{\intercal}M^{-1}x_2=x_2^{\intercal}M^{-1}x_1$ by symmetry.
	\end{proof}
	
	\begin{remark}
		For a right extension, take $M=A_{\pi_1}+A_{\pi_2}$, $x_i=e_{\pi_i(n)}$, and $y_i=e_{\rho_i(1)}$ in Lemma~\ref{lem:offset}, then \eqref{Eq_offset}  corresponds to 
		the matrix $(V_1+V_2)^\intercal (A_{\pi_1}+ A_{\pi_2})^{-1}(V_1+V_2)$ in the Schur complement in Lemma~\ref{thm_comp_extension_mat}.
		We call it the \textit{offset matrix} in the Schur complement.
		In this case, the offset matrix
		is given by
		$$
		(x_1^\intercal M^{-1}x_2)\Delta, \text{ where } \Delta=y_1y_2^\intercal+y_2y_1^\intercal \in \F_2^{t\times t}.
		$$
		If the blocks start at the same vertex, i.e., $\rho_1(1)=\rho_2(1)$, then $y_1=y_2$ and $\Delta = \mathbf{0}_{t\times t}$. This explains the unchanged Schur complement in Example~\ref{ex:four-block-repeat}.
	\end{remark}
	
	For iterative right-extensions, we give a simplified characterization on the compatibility of two extension permutations,
	which facilitate the construction of compatible sets with mixed extensions.
	
	\begin{proposition}\label{prop3}
		Let $n,t$ be even, let $\pi_1,\pi_2\in S_n$ be compatible and $\rho_1,\rho_2\in S_t$ be compatible, and let $A=A_{\pi_1}+A_{\pi_2}$ and $B=A_{\rho_1}+A_{\rho_2}$. Let
		\[
		\vartheta_i^{(0)}=\pi_i,\quad
		\vartheta_i^{(k)}=\pi_i\|(\rho_i^{(k)}+n)\in S_{n+tk},\quad i=1,2,\quad k\geq1,
		\]
		and denote
		\begin{equation}\label{eq:abc}
			a=(A^{-1})_{\pi_1(n),\pi_2(n)},\quad
			b=(B^{-1})_{\rho_1(1),\rho_2(1)},\quad
			c=(B^{-1})_{\rho_1(t),\rho_2(t)}.
		\end{equation}
		Then $\vartheta_1^{(k)}$ and $\vartheta_2^{(k)}$ are compatible for all $k\geq1$ if and only if $ab=bc=0$.
	\end{proposition}
	
	\begin{proof}
		For $k\geq0$, let $M_k=A_{\vartheta_1^{(k)}}+A_{\vartheta_2^{(k)}}$ and $\ell_k=n+tk$. When $M_k$ is nonsingular, let
		\[
		a_k=\bigl(M_k^{-1}\bigr)_{\vartheta_1^{(k)}(\ell_k),\,\vartheta_2^{(k)}(\ell_k)}
		\]
		be the entry of $M_k^{-1}$ indexed by the right endpoints of $\vartheta_1^{(k)}$ and $\vartheta_2^{(k)}$. In particular, $a_0=a$. 
		Suppose $k\ge1$ and the two paths at stage $k-1$ are compatible. We will show that (i) the paths at stage $k$ are compatible if and only if $a_{k-1}b=0$, and (ii) in that case $a_k=c+(1+c)a_{k-1}$.
		
		\smallskip
		
		For simplicity we denote $u_i=e_{\rho_i(1)}$ for $i=1,2$, and take $U=[u_1\ u_2]\in\F_2^{t\times2}$ and $J=\begin{bsmallmatrix}0&1\\1&0\end{bsmallmatrix}$. Then $\Delta=u_1u_2^{\intercal}+u_2u_1^{\intercal}=UJU^{\intercal}$. 
		Suppose that for $k\geq1$ the permutations $\vartheta_1^{(k-1)}$ and $\vartheta_2^{(k-1)}$ are compatible, i.e., $M_{k-1}$ is nonsingular.
		
		We first consider the Schur complement. The permutation $\vartheta_i^{(k)}$ is the extension of $\vartheta_i^{(k-1)}$ by $\rho_i$ at position $0$. By Lemma~\ref{lem_adjmat},
		\[
		M_k=\begin{bmatrix}M_{k-1}&V^{(k-1)}\\(V^{(k-1)})^{\intercal}&B\end{bmatrix},\qquad
		V^{(k-1)}=e_{\vartheta_1^{(k-1)}(\ell_{k-1})}u_1^{\intercal}+e_{\vartheta_2^{(k-1)}(\ell_{k-1})}u_2^{\intercal}.
		\]
		Applying Lemma~\ref{lem:offset} with $M=M_{k-1}$, $x_i=e_{\vartheta_i^{(k-1)}(\ell_{k-1})}$ and $y_i=u_i$ gives $$(V^{(k-1)})^{\intercal}M_{k-1}^{-1}V^{(k-1)}=a_{k-1}(u_1u_2^{\intercal}+u_2u_1^{\intercal})=a_{k-1}\Delta.$$ Hence the Schur complement of $M_{k-1}$ in $M_k$ is given by
		\begin{equation*}
			C_k=B+a_{k-1}\Delta,
		\end{equation*}
		and by Lemma~\ref{thm_comp_extension_mat}, $\vartheta_1^{(k)}$ and $\vartheta_2^{(k)}$ are compatible if and only if $C_k$ is nonsingular. 
		
		\smallskip
		
		If $a_{k-1}=0$, then $C_k=B$ is nonsingular immediately. Suppose $a_{k-1}=1$. Then  $C_k=B+\Delta=B+UJU^{\intercal}$. Consider the following two block matrices.
		\[
		H=\begin{bmatrix}I_2&JU^{\intercal}\\U&B\end{bmatrix}, 
		H' = \begin{bmatrix}B&U\\JU^{\intercal}&I_2\end{bmatrix}
		\in\F_2^{(t+2)\times(t+2)}.
		\]
		Interchanging the two block rows and the corresponding block columns takes $H$ to $H'$ and preserves the determinant. 
		Applying Lemma~\ref{lem_SC} to the block $I_2$ in $H$ gives the equality $\det(H)=\det(B+UJU^{\intercal})=\det(C_k)$.
		
		Applying Lemma~\ref{lem_SC} to $H'$ with respect to $B$ gives 
		\begin{align*}
			\det(C_k)&=\det(H')=\det(B)\det\bigl(I_2+JU^{\intercal}B^{-1}U\bigr)\\
			&=\det(I_2+bJ^2)=\det((1+b)I_2)=1+b.
		\end{align*}
		Therefore, in both cases $C_k$ is nonsingular if and only if $a_{k-1}b=0$. This proves (i).
		
		Furthermore, when $C_k$ is nonsingular, 
		by~\eqref{eq:block-inverse}, the lower-right $t\times t$ block of $M_k^{-1}$ is $C_k^{-1}$. The right endpoints of $\vartheta_1^{(k)}$ and $\vartheta_2^{(k)}$ are $\ell_{k-1}+\rho_1(t)$ and $\ell_{k-1}+\rho_2(t)$, which correspond to the rows and columns $\rho_1(t),\rho_2(t)$ of this block. Thus, letting $w_i=e_{\rho_i(t)}$ in $\F_2^t$ we have
		$a_k=w_1^{\intercal}C_k^{-1}w_2$. 
		
		If $a_{k-1}=0$, then $C_k=B$ and $a_k=c$. Suppose $a_{k-1}=1$. By (i) we have $b=0$, and hence $U^{\intercal}B^{-1}U=\mathbf 0_{2\times2}$. A direct multiplication gives
		\[
		(B+UJU^{\intercal})(B^{-1}+B^{-1}UJU^{\intercal}B^{-1})
		=I_t+UJ\bigl(U^{\intercal}B^{-1}U\bigr)JU^{\intercal}B^{-1}=I_t,
		\]
		implying that 
		$C_k^{-1}=B^{-1}+B^{-1}UJU^{\intercal}B^{-1}.$
		Since $UJU^{\intercal}=u_1u_2^{\intercal}+u_2u_1^{\intercal}$, we have
		\(
		a_k=w_1^{\intercal}B^{-1}w_2+w_1^{\intercal}B^{-1}\bigl(u_1u_2^{\intercal}+u_2u_1^{\intercal}\bigr)B^{-1}w_2.
		\)
		To evaluate the correction term, let $d_{ij}=u_i^{\intercal}B^{-1}w_j$. Each path has odd degree precisely at its two endpoints, so $B\mathbf 1=u_1+u_2+w_1+w_2$ over $\Ftwo$. Multiplying this identity on the left by $u_i^{\intercal}B^{-1}$ gives $d_{i1}+d_{i2}=1$, since $U^{\intercal}B^{-1}U=\mathbf{0}_{2
        \times 2}$. Multiplication by $w_1^{\intercal}B^{-1}$ gives $d_{11}+d_{21}+c=1$, since $B^{-1}$ is alternating. Thus the correction term is
		\[
		d_{11}d_{22}+d_{21}d_{12}
		=d_{11}(1+d_{21})+d_{21}(1+d_{11})
		=d_{11}+d_{21}=1+c.
		\]
		It follows that $a_k=c+(1+c)=1$ when $a_{k-1}=1$.
		Combining both cases gives the expression $a_k=c+(1+c)a_{k-1}$ as in (ii). 
		
		\medskip
		
		The recurrence starts with the nonsingular matrix $M_0=A$ and $a_0=a$. If $b=0$, the product $a_{k-1}b$ vanishes at every stage where $a_{k-1}$ is defined. Part (i) then gives nonsingularity of $M_k$ by induction, starting at $M_0$.
		
		If $b=1$, part (i) requires $a=0$ for the first extension to be compatible. When $a=0$, part (ii) gives $a_1=c$. Hence compatibility at the second extension requires $c=0$. Conversely, if $a=c=0$, part (ii) gives $a_k=0$ after every compatible extension, and part (i) yields compatibility at the next stage. These two cases show that the entire sequence is compatible exactly when $ab=bc=0$.
	\end{proof}
	
	\begin{theorem}
		\label{Th_mix_extension}
		Let $n,t$ be even. Let $S = \{\pi_1,\dots, \pi_L\}\subset S_n$ and $T = \{\rho_1,\dots, \rho_L\}\subset S_t$ be two compatible sets, and for any $1\leq i<j\leq L$,
		$a_{ij},b_{ij},c_{ij}$ be the entries~\eqref{eq:abc} for the pairs $(\pi_i,\pi_j)$ and $(\rho_i,\rho_j)$. Then
		\(
		\bigl\{\pi_i\,\|\,(\rho_i^{(k)}+n):\ 1\le i\le L\bigr\}
		\)
		is a compatible set of size $L$ in $S_{n+tk}$ for every $k\ge0$ if and only if
		\begin{equation}\label{eq:test}
			a_{ij}b_{ij}=b_{ij}c_{ij}=0\qquad\text{for all } 1\le i<j\le L.
		\end{equation}
	\end{theorem}
	\begin{proof}
		The extended paths are distinct because their first $n$ entries recover the seeds. The family is compatible at every iteration if and only if each pair is compatible at every iteration. Applying Proposition~\ref{prop3} to each pair gives precisely~\eqref{eq:test}.
	\end{proof}
	
	\section{Compatible sets from the Schur-complement criterion}
	\label{sec:constructions}
	
	The three-entry criterion applies both when a seed is repeated and when it is extended by a different block. We first determine how the criterion simplifies for self-extensions.
	
	\subsection{Compatible sets under self-extension}
	
	When each seed is its own block, the initial and final right-endpoint entries coincide. The three-entry test then reduces to a single product for each pair.
	\begin{theorem}\label{Th-CS-self-extension}
		Let $t$ be even and let $S=\{\pi_1,\ldots,\pi_L\}\subseteq S_t$ be a compatible set. For $1\le i<j\le L$, set 
		\[
		A=(A_{\pi_i}+A_{\pi_j})^{-1},
		a_{ij}=A_{\pi_i(t),\pi_j(t)}, \ b_{ij}=A_{\pi_i(1),\,\pi_j(1)}.
		\]
		Then $\{\pi^{(k)}:\pi\in S\}$ is a compatible set of size $L$ in $S_{tk}$ for every $k\ge1$ if and only if
		\begin{equation}\label{eq:cond_se}
			a_{ij}b_{ij}=0 \text{ for all } (i,j) \text{ with } 1\leq i<j\leq L.
		\end{equation} 
	\end{theorem}
	\begin{proof}
		For each $i$, the first $t$ entries of $\pi_i^{(k)}$ recover $\pi_i$. Thus the self-extended family has size $L$ for every $k\ge1$. For a fixed pair $i<j$, we take $\pi_i,\pi_j$ both as the seeds and as the length-$t$ blocks. Appending $\ell$ translated copies to these seeds gives $\pi_i^{(\ell+1)}$ and $\pi_j^{(\ell+1)}$, since $\pi_i\|(\pi_i^{(\ell)}+t)=\pi_i^{(\ell+1)}$ and the same identity holds for $\pi_j$.
		
		The seed and block adjacency sums are both $A_{\pi_i}+A_{\pi_j}$. In~\eqref{eq:abc}, the entry $a$ comes from the last endpoints of the seeds, while $c$ comes from the last endpoints of the blocks. These endpoints agree, so $a=c=a_{ij}$. The entry $b$ comes from the first endpoints of the blocks and equals $b_{ij}$.
		
		Proposition~\ref{prop3} therefore gives compatibility for this pair at every $\ell\ge1$ if and only if both $ab$ and $bc$ vanish. Since both products equal $a_{ij}b_{ij}$, this is precisely the condition in~\eqref{eq:cond_se} for the pair $i<j$. At $\ell=0$ the family is the compatible seed set $S$. Taking all pairs proves the claimed equivalence.
	\end{proof}
	
We call a compatible set \emph{self-extendable} if it satisfies~\eqref{eq:cond_se}. The next two corollaries give explicit self-extendable sets in dimensions  $4$ and $6$.
We first recover the dimension-$4$ construction by repeating six compatible paths~\cite{CCDS25}.
	\begin{corollary}\label{cor-ccds25}
		The set
		\[
		S=\{[1,2,3,4],[1,3,4,2],[1,4,2,3],[2,1,4,3],[2,3,1,4],[3,1,2,4]\}
		\]
		is self-extendable. Its $k$-fold self-extensions form a compatible set of size six in dimension $4k$ for every $k\ge1$.
	\end{corollary}
	\begin{proof}
		We index the six permutations as $\pi_1,\ldots,\pi_6$ in the displayed order and write $A_{ij}=A_{\pi_i}+A_{\pi_j}$. The entry at $(u,v)$ of $A_{ij}$ is one precisely when the edge $\{u,v\}$ occurs in exactly one of the two paths. For instance, $\pi_1$ and $\pi_2$ share the edge $\{3,4\}$. The edges of $A_{12}$ are $\{1,2\}$, $\{2,3\}$, $\{1,3\}$, and $\{2,4\}$. They give
		\[
		A_{12}=
		\begin{pmatrix}
			0&1&1&0\\
			1&0&1&1\\
			1&1&0&0\\
			0&1&0&0
		\end{pmatrix},
		\qquad
		(A_{12})^{-1}=
		\begin{pmatrix}
			0&0&1&1\\
			0&0&0&1\\
			1&0&0&1\\
			1&1&1&0
		\end{pmatrix}.
		\]
		The first endpoints of $\pi_1,\pi_2$ are both one, so $b_{12}=(A_{12}^{-1})_{1,1}=0$. The last endpoints are four and two, so $a_{12}=(A_{12}^{-1})_{4,2}=1$.
		
		For all $15$ pairs, row reduction of $[A_{ij}\mid I_4]$ gives four pivots in the left block. The table lists the entries $(a_{ij},b_{ij})$ of the inverse in the right block.
		\[
		\begin{array}{c|ccccc}
			i\backslash j&2&3&4&5&6\\\hline
			1&(1,0)&(1,0)&(0,0)&(0,1)&(0,1)\\
			2&     &(1,0)&(1,0)&(0,1)&(0,0)\\
			3&     &     &(0,1)&(0,0)&(1,0)\\
			4&     &     &     &(1,0)&(0,1)\\
			5&     &     &     &     &(0,1)
		\end{array}
		\]
		Nonsingularity gives compatibility of the seed set. Every entry in the table has product zero, so Theorem~\ref{Th-CS-self-extension} gives compatibility after every repetition. The six repeated paths are distinct because they have different first blocks.
	\end{proof}
	
	The four-vertex construction repeats six paths. Two six-vertex seed families satisfy the same condition with $8$ paths each.
	
	\begin{corollary}\label{t=6}
		The $6$-dimensional sets
		\begin{equation*}
			\begin{split}
				S_1 =
				& \{ [1, 2, 3, 4, 5, 6],
				[1, 5, 3, 2, 6, 4],
				[2, 1, 6, 4, 3, 5],
				[5, 4, 2, 1, 3, 6],
				\\ &  \,\,\,	
				[1, 4, 5, 3, 6, 2],
				[2, 5, 6, 1, 4, 3],
				[4, 2, 3, 6, 1, 5],
				[3, 2, 5, 1, 4, 6]\},\\
				S_2 = 
				& 
				\{
				[1, 2, 3, 4, 5, 6],
				[1, 3, 6, 2, 5, 4],
				[4, 3, 6, 1, 2, 5],
				[1, 4, 6, 5, 3, 2],
				\\&
				\,\,\,[3, 1, 5, 4, 2, 6],
				[2, 6, 1, 4, 5, 3],
				[2, 4, 3, 1, 6, 5],
				[5, 1, 4, 2, 3, 6]
				\} 
			\end{split}
		\end{equation*}
		are self-extendable. For $i=1,2$ and $k\ge1$, the set $\{\pi^{(k)}:\pi\in S_i\}$ is compatible and has size $8$.
		
	\end{corollary}
	\begin{proof}
		Within each of the two displayed families, we number the paths from one to $8$. For each $i<j$, we form $D_{ij}=A_{\pi_i}+A_{\pi_j}$ and row-reduce $[D_{ij}\mid I_6]$ over $\Ftwo$. In $S_1$, the first pair illustrates the endpoint calculation. Its adjacency sum has edges
		\[
		\{1,2\},\ \{1,5\},\ \{2,6\},\ \{3,4\},\
		\{3,5\},\ \{4,5\},\ \{4,6\},\ \{5,6\}.
		\]
		Its first endpoints agree, so $b_{12}=0$. The vectors are indexed by vertex labels, and the solution of $D_{12}x=e_4$ is $x=(1,0,0,0,0,1)^\intercal$. Since the last endpoints are six and four, the last-endpoint entry is $a_{12}=x_6=1$.
		
		All $28$ adjacency sums in each family have six pivots in the left block. For the endpoint test it suffices to list the pairs with $b_{ij}=1$. The row reductions give the following lists.
		\[
		\begin{aligned}
			S_1\quad & (1,3),(1,4),(1,8),(2,7),(3,5),(3,7),\\[-2pt]
			& (4,7),(4,8),(5,7),(6,7),(6,8),(7,8),\\
			S_2\quad & (1,5),(1,6),(1,8),(3,7),(3,8),(4,7),(4,8),(5,7),(5,8).
		\end{aligned}
		\]
		For every pair in the lists the inverse entry at the last endpoints is zero. Every other pair has $b_{ij}=0$. Thus each family is compatible and satisfies~\eqref{eq:cond_se}. Theorem~\ref{Th-CS-self-extension} now gives the stated $8$ paths in each dimension $6k$.
	\end{proof}
	% Increasing the block length to $8$ gives a family of eleven paths whose translated copies can be appended indefinitely. 
	
	% \begin{corollary} \label{t=8}
		% Let 
		% 	\begin{equation*}
			% 		\begin{split}
				% S=\{ & [1, 2, 3, 4, 5, 6, 7, 8],[1, 4, 2, 8, 3, 7, 5, 6],[1, 3, 2, 6, 8, 4, 5, 7],[1, 5, 7, 2, 8, 4, 6, 3], \\
				% 			& [2, 8, 5, 7, 1, 3, 4, 6],[2, 5, 4, 1, 6, 8, 7, 3],[3, 2, 4, 7, 1, 6, 5, 8],[4, 3, 5, 2, 6, 1, 8, 7], \\& [6, 3, 5, 4, 2, 7, 1, 8],[6, 1, 5, 2, 3, 8, 4, 7],[7, 3, 1, 5, 4, 6, 2, 8]\}.
				% 		\end{split}
			% 	\end{equation*}
		% Then $\{\pi^{(k)}:  \pi \in S\}$ is a compatible set of size $11$ in dimension $8k$ for any $k\geq 1$.
		% \end{corollary}
	% \begin{proof}
		% For the $55$ pairs in the displayed seed, Gaussian elimination over $\Ftwo$ gives a nonsingular adjacency sum in every case. The first-endpoint entry $b_{ij}$ equals one for 20 pairs. For each of these pairs the last-endpoint entry $a_{ij}$ is zero. Theorem~\ref{Th-CS-self-extension} now applies for every $k\ge1$.
		% \end{proof}
	
	In dimension six, ordinary compatibility and compatibility under repeated self-extension give different maxima.
	
	\begin{proposition}\label{prop:sixexact}
		One has $\Mb(6)=9$. The maximum size of a self-extendable compatible set in $S_6$ is $8$.
	\end{proposition}
	\begin{proof}
		Simultaneous relabeling preserves compatibility and conjugates the inverse adjacency sums by the same permutation matrix, so it also preserves the self-extension test. We may thus require the identity path in each family. For ordinary compatibility we keep only the orientation with the smaller first endpoint, since reversal does not change the adjacency matrix. For self-extension we keep both orientations, since reversing one path can change the endpoint test. We make two candidates adjacent when their adjacency sum has rank six and, for self-extension, their inverse endpoint entries satisfy~\eqref{eq:cond_se}.
		
		Order the candidates lexicographically and start with the identity and its neighbors. At a node with $r$ selected paths, count one family of size $r$. For each remaining candidate $v$, recurse with $v$ selected and retain only its neighbors greater than $v$ among the remaining candidates. Every family containing the identity is counted once because the other paths are added in increasing order. This gives the following counts.
		\[
		\begin{array}{crrrrrrrrr}
           \toprule
			r&1&2&3&4&5&6&7&8&9\\ \midrule
			\text{ordinary}&1&128&2920&15624&21252&8316&1484&336&42\\
			\text{self-extension}&1&228&8180&59648&92610&34344&4368&512&0 \\
            \bottomrule
		\end{array}
		\]
		In the ordinary case every node of size nine has an empty remaining candidate set. Hence the maxima are nine and $8$, respectively.
	\end{proof}
	
	\subsection{Compatible sets under mixed extensions}\label{subsec_mix_right_ext}
	
	%The self-extension condition constrains the first and last endpoints of the same path pair. The mixed criterion allows us to choose the seeds separately from the blocks. For mixed extensions, we choose even dimensions $n,t$ and compatible blocks $\rho_1,\dots,\rho_L\in S_t$ satisfying $b_{ij}c_{ij}=0$. We then seek compatible seeds $\pi_1,\dots,\pi_L\in S_n$ with $a_{ij}=0$ whenever $b_{ij}=1$. The first example uses seeds in dimension six and blocks in dimension four.
	
	To construct compatible sets in $S_{n+tk}$, we choose even integers $n\ge t$ and apply Theorem~\ref{Th_mix_extension} in two stages. We first choose a self-extendable compatible family $(\rho_1,\ldots,\rho_L)$ in $S_t$, whose members serve as extension blocks. By Theorem~\ref{Th-CS-self-extension}, the block entries in~\eqref{eq:abc} satisfy $b_{ij}c_{ij}=0$ for every $i<j$. We then seek a compatible seed family $(\pi_1,\ldots,\pi_L)$ in $S_n$, with $\pi_i$ assigned to $\rho_i$, such that $a_{ij}=0$ for each pair with $b_{ij}=1$. The other pairs satisfy $a_{ij}b_{ij}=0$ because $b_{ij}=0$. Thus the matched families give $L$ compatible permutations in $S_{n+tk}$ for every $k\ge0$.
	
	For $n=6$ and $t=4$, a direct search yields the matching in the next corollary. For the larger families considered afterward, cyclic symmetry reduces the number of pairwise checks.

	\begin{corollary}\label{cor-4k+2}
		Let $S=(\pi_1,\ldots,\pi_6)$ and $T=(\rho_1,\ldots,\rho_6)$ be the following ordered families:
		\[
		\begin{aligned}
			S={}&([4,1,3,2,6,5],[1,3,6,5,2,4],[1,5,4,2,6,3],\\
			&\phantom{(}[3,4,6,2,1,5],[2,3,5,4,1,6],[2,6,1,5,3,4]),\\
			T={}&([1,2,3,4],[3,1,2,4],[1,4,2,3],[2,3,1,4],[1,3,4,2],[2,1,4,3]).
		\end{aligned}
		\]
		Then the set $\{\pi_i\|(\rho_i^{(k)}+6): i =1,\dots, 6\}$ is a compatible set in dimension $4(k+1)+2$ for any $k\geq 1$.
	\end{corollary}
	\begin{proof}
		Row reduction over $\Ftwo$ gives rank six for the $15$ seed adjacency sums and rank four for the $15$ block adjacency sums. Thus both families are compatible. For each pair we compute the entries in~\eqref{eq:abc}. It suffices to show that for every pair $i<j$, either $b_{ij}=0$, or $b_{ij}=1$ and $a_{ij}=c_{ij}=0$.
		
		For $(i,j)=(1,2)$ the seed paths and blocks are
		\[
		\pi_1=[4,1,3,2,6,5],\quad \pi_2=[1,3,6,5,2,4],\qquad
		\rho_1=[1,2,3,4],\quad \rho_2=[3,1,2,4].
		\]
		In $A=A_{\pi_1}+A_{\pi_2}$ the common edges $(1,3)$ and $(5,6)$ cancel, leaving the edges $(1,4),(2,3),(2,4),(2,5),(2,6),(3,6)$. In $B=A_{\rho_1}+A_{\rho_2}$ the common edge $(1,2)$ cancels, leaving the edges $(1,3),(2,3),(2,4),(3,4)$. Thus we have
		\[A=\begin{pmatrix}
			0&0&0&1&0&0\\ 0&0&1&1&1&1\\ 0&1&0&0&0&1\\
			1&1&0&0&0&0\\ 0&1&0&0&0&0\\ 0&1&1&0&0&0
		\end{pmatrix},\
		A^{-1}=\begin{pmatrix}
			0&0&0&1&1&0\\ 0&0&0&0&1&0\\ 0&0&0&0&1&1\\
			1&0&0&0&0&0\\ 1&1&1&0&0&1\\ 0&0&1&0&1&0
		\end{pmatrix},
		\] and
		\[
		B=\begin{pmatrix}0&0&1&0\\0&0&1&1\\1&1&0&1\\0&1&1&0\end{pmatrix},\
		B^{-1}=\begin{pmatrix}0&1&1&1\\1&0&0&1\\1&0&0&0\\1&1&0&0\end{pmatrix}.
		\]
		The relevant endpoints are $\pi_1(6)=5$, $\pi_2(6)=4$, $\rho_1(1)=1$, $\rho_2(1)=3$ and $\rho_1(4)=\rho_2(4)=4$. Hence
		\[
		a_{12}=(A^{-1})_{5,4}=0,\qquad b_{12}=(B^{-1})_{1,3}=1,\qquad c_{12}=(B^{-1})_{4,4}=0,
		\]
		where $c_{12}=0$ also follows from Lemma~\ref{lem:offset}, since $B^{-1}$ has zero diagonal. Since $b_{12}=1$ and $a_{12}=c_{12}=0$, the permutations $\pi_1\|(\rho_1^{(k)}+6)$ and $\pi_2\|(\rho_2^{(k)}+6)$ are compatible for $k\geq1$ by Proposition~\ref{prop3}.
		Applying the same row reductions to the remaining $14$ pairs gives the triples $(a_{ij},b_{ij},c_{ij})$ in the following table.
		\begin{center}\small
			\setlength{\tabcolsep}{4pt}
			\begin{tabular}{|@{}c|ccccccc@{}|}\hline
				$(i,j)$ & $(1,3)$ & $(1,4)$ & $(1,5)$ & $(1,6)$ & $(2,3)$ & $(2,4)$ & $(2,5)$\\
				$(a_{ij},b_{ij},c_{ij})$ & $(0,0,1)$ & $(0,1,0)$ & $(0,0,1)$ & $(1,0,0)$ & $(1,0,1)$ & $(0,1,0)$ & $(1,0,0)$\\ \hline
				$(i,j)$ & $(2,6)$ & $(3,4)$ & $(3,5)$ & $(3,6)$ & $(4,5)$ & $(4,6)$ & $(5,6)$\\
				$(a_{ij},b_{ij},c_{ij})$ & $(0,1,0)$ & $(1,0,0)$ & $(1,0,1)$ & $(0,1,0)$ & $(0,1,0)$ & $(0,0,1)$ & $(1,0,1)$\\
				\hline
			\end{tabular}
		\end{center}
		Since every pair with $b_{ij}=1$, namely $(1,2),(1,4),(2,4),(2,6),(3,6),(4,5)$, has $a_{ij}=c_{ij}=0$, all $15$ pairs of the extended permutations are compatible for any $k\geq1$. Thus the extended set has size $6$.
	\end{proof}
	
	The preceding corollary uses a direct search in dimensions six and four. For larger families, a union of cyclic orbits reduces the number of pairs to be tested. For $\omega\in S_t$ of order $m$, we write $\omega(\rho)=\omega\circ\rho$ for entrywise relabeling of a path $\rho\in S_t$. The orbit of $\rho$ under $\Omega=\langle\omega\rangle$ is $\Omega(\rho)=\{\omega^k(\rho):0\le k<m\}$. If $\omega^k(\rho)=\rho$, then $\omega^k$ fixes every label since $\rho$ is a permutation. Thus each orbit has $m$ elements.
	
	Recall that for compatible seeds $\pi_1,\pi_2\in S_n$ and compatible blocks $\rho_1,\rho_2\in S_t$, Proposition~\ref{prop3} gives compatibility of $\pi_1\|(\rho_1^{(k)}+n)$ and $\pi_2\|(\rho_2^{(k)}+n)$ for every $k\ge1$ if and only if $ab=bc=0$, with $a,b,c$ defined in~\eqref{eq:abc}. We first reduce the pairwise tests for self-extensions, where each path serves as both seed and block.
	
	For even $t$, we call a pair $(\rho,\sigma)$ in $S_t$ \emph{admissible} if $\rho\sim\sigma$ and $ab=0$, where
	\[
	a=\bigl((A_\rho+A_\sigma)^{-1}\bigr)_{\rho(t),\sigma(t)},
	\qquad
	b=\bigl((A_\rho+A_\sigma)^{-1}\bigr)_{\rho(1),\sigma(1)}.
	\]
	The entries are defined for compatible pairs, for which the adjacency sum is nonsingular.
	
	\begin{lemma}\label{lem:relabel-invariance}
		Let $\rho,\sigma,\omega\in S_t$. Then $\rho\sim\sigma$ if and only if $\omega(\rho)\sim\omega(\sigma)$. If $A_\rho+A_\sigma$ is nonsingular, the first- and last-endpoint entries of its inverse agree with the corresponding entries for the relabeled pair. In particular, for even $t$, $(\rho,\sigma)$ is admissible if and only if $(\omega(\rho),\omega(\sigma))$ is admissible.
	\end{lemma}
	\begin{proof}
		For a relabeling $\omega$, let $P_\omega$ be its permutation matrix, with $P_\omega e_i=e_{\omega(i)}$. Then $A_{\omega(\rho)}+A_{\omega(\sigma)}=P_\omega(A_\rho+A_\sigma)P_\omega^\intercal$, so the ranks agree. If the sum is nonsingular, its inverse transforms by the same conjugation. The endpoints are relabeled by $\omega$, so their corresponding inverse entries agree.
	\end{proof}
	
	\begin{proposition}\label{prop:orbit-reduction}
		Let $t$ be even and let $\Omega=\langle\omega\rangle\le S_t$ have order $m$. Let $\rho_1,\ldots,\rho_r\in S_t$ represent distinct $\Omega$-orbits and set $U=\bigcup_{i=1}^r\Omega(\rho_i)$. Then $U$ is self-extendable if and only if the following conditions hold.
		\begin{enumerate}[label=$(\roman*)$]
			\item For comparisons within an orbit, $(\rho_i,\omega^s(\rho_i))$ is admissible for $1\le i\le r$ and $1\le s\le\lfloor m/2\rfloor$.
			\item For comparisons between distinct orbits, $(\rho_i,\omega^s(\rho_j))$ is admissible for $1\le i<j\le r$ and $0\le s<m$.
		\end{enumerate}
		In this case $U$ is a self-extendable compatible set of size $rm$.
	\end{proposition}
	\begin{proof}
		Each distinct pair in $U$ has the form $(\omega^k(\rho_i),\omega^\ell(\rho_j))$. Relabeling both members by $\omega^{-k}$ gives $(\rho_i,\omega^s(\rho_j))$, where $s=\ell-k\pmod m$. Lemma~\ref{lem:relabel-invariance} shows that this relabeling preserves admissibility. When $i\ne j$, we can order the two orbits so that $i<j$, and the relative exponent may be any residue modulo $m$. When $i=j$, distinct paths have $s\ne0$. Interchanging the paths and relabeling gives $(\rho_i,\omega^{-s}(\rho_i))$, so the shifts $s$ and $-s$ give the same test. This gives the two ranges in the statement.
		
		Both ranges consist of comparisons between distinct members of $U$. Thus the conditions are equivalent to admissibility of every distinct pair, which is equivalent to self-extendability by Theorem~\ref{Th-CS-self-extension}. The orbits are disjoint by hypothesis and each contains $m$ paths, so $|U|=rm$.
	\end{proof}
	
	Tables~\ref{tab:self_ext_8} and~\ref{tab:self_ext_10} give self-extendable compatible sets of sizes $15$ and $18$. Using the group cycle notation, which shows how each element maps to the next, wrapping back to the start, the relabelings are $\omega_8=(1\,2\,3\,4\,5)$ and $\omega_{10}=(1\,2\,3\,4\,5\,6\,7\,8\,9)$, with all unlisted labels fixed. In each table the first row gives representatives of distinct orbits. Gaussian elimination over $\Ftwo$ verifies nonsingularity in all $21$ reduced comparisons for $t=8$ and all $17$ for $t=10$. In these comparisons the first-endpoint entry is one in $8$ cases for $t=8$ and four for $t=10$, and the last-endpoint entry is zero in each such case. Proposition~\ref{prop:orbit-reduction} therefore applies.
	
	\begin{table}[t]
		\centering\small
		\caption{A self-extendable compatible set of size $15$ in $S_8$}\label{tab:self_ext_8}
		\begin{tabular}{@{}cccc@{}}
			\toprule
			$k$ & $\omega^k(\rho_1)$ & $\omega^k(\rho_2)$ & $\omega^k(\rho_3)$\\
			\midrule
			$0$ & $[8,1,5,7,4,6,3,2]$ & $[1,7,4,2,6,5,8,3]$ & $[1,6,7,4,8,3,5,2]$\\
			$1$ & $[8,2,1,7,5,6,4,3]$ & $[2,7,5,3,6,1,8,4]$ & $[2,6,7,5,8,4,1,3]$\\
			$2$ & $[8,3,2,7,1,6,5,4]$ & $[3,7,1,4,6,2,8,5]$ & $[3,6,7,1,8,5,2,4]$\\
			$3$ & $[8,4,3,7,2,6,1,5]$ & $[4,7,2,5,6,3,8,1]$ & $[4,6,7,2,8,1,3,5]$\\
			$4$ & $[8,5,4,7,3,6,2,1]$ & $[5,7,3,1,6,4,8,2]$ & $[5,6,7,3,8,2,4,1]$\\
			\bottomrule
		\end{tabular}
	\end{table}
	
	\begin{table}[t]
		\centering\small
		\caption{A self-extendable compatible set of size $18$ in $S_{10}$}\label{tab:self_ext_10}
		\begin{tabular}{@{}ccc@{}}
			\toprule
			$k$ & $\omega^k(\rho_1)$ & $\omega^k(\rho_2)$\\
			\midrule
			$0$ & $[10,5,7,6,8,4,9,3,2,1]$ & $[8,4,10,9,7,6,3,2,5,1]$\\
			$1$ & $[10,6,8,7,9,5,1,4,3,2]$ & $[9,5,10,1,8,7,4,3,6,2]$\\
			$2$ & $[10,7,9,8,1,6,2,5,4,3]$ & $[1,6,10,2,9,8,5,4,7,3]$\\
			$3$ & $[10,8,1,9,2,7,3,6,5,4]$ & $[2,7,10,3,1,9,6,5,8,4]$\\
			$4$ & $[10,9,2,1,3,8,4,7,6,5]$ & $[3,8,10,4,2,1,7,6,9,5]$\\
			$5$ & $[10,1,3,2,4,9,5,8,7,6]$ & $[4,9,10,5,3,2,8,7,1,6]$\\
			$6$ & $[10,2,4,3,5,1,6,9,8,7]$ & $[5,1,10,6,4,3,9,8,2,7]$\\
			$7$ & $[10,3,5,4,6,2,7,1,9,8]$ & $[6,2,10,7,5,4,1,9,3,8]$\\
			$8$ & $[10,4,6,5,7,3,8,2,1,9]$ & $[7,3,10,8,6,5,2,1,4,9]$\\
			\bottomrule
		\end{tabular}
	\end{table}

	The same relabeling argument applies when each seed is assigned a separate block. We write $a(\pi,\pi')$, $b(\rho,\rho')$, and $c(\rho,\rho')$ for the entries in~\eqref{eq:abc} to indicate their dependence on the seed or block pair.
	
	\begin{proposition}\label{prop:orbit-matching}
		Let $d,t$ be even and let $\theta\in S_d$ and $\omega\in S_t$ have the same order $m$. Suppose $\pi_1,\ldots,\pi_r\in S_d$ represent distinct $\langle\theta\rangle$-orbits and $\rho_1,\ldots,\rho_r\in S_t$ represent distinct $\langle\omega\rangle$-orbits. We write
		\[
		\pi_{i,k}=\theta^k(\pi_i),\qquad
		\rho_{i,k}=\omega^k(\rho_i)
		\qquad (1\le i\le r,\ 0\le k<m).
		\]
		We match $\pi_{i,k}$ with $\rho_{i,k}$. The seed and block families are compatible, and
		\[
		\bigl\{\pi_{i,k}\,\|\,(\rho_{i,k}^{(h)}+d)
		\mid 1\le i\le r,\ 0\le k<m\bigr\}
		\]
		is a compatible set of size $rm$ in $S_{d+th}$ for every $h\ge0$ if and only if the following tests hold for every triple $(i,j,s)$ in Proposition~\textup{\ref{prop:orbit-reduction}}\textup{(i)} or \textup{(ii)}. With $\pi=\pi_{i,0}$, $\pi'=\pi_{j,s}$, $\rho=\rho_{i,0}$ and $\rho'=\rho_{j,s}$, the tests are
		\[
		\pi\sim\pi',\qquad
		\rho\sim\rho',\qquad
		a(\pi,\pi')\,b(\rho,\rho')=0,\qquad
		b(\rho,\rho')\,c(\rho,\rho')=0.
		\]
	\end{proposition}
	\begin{proof}
		The distinct orbits and the fact that each orbit has $m$ elements give $rm$ distinct seeds and $rm$ distinct blocks. The matching pairs them by their common indices. For two indices $(i,k)$ and $(j,\ell)$, we relabel both seeds by $\theta^{-k}$ and both blocks by $\omega^{-k}$. The resulting comparison is between the indices $(i,0)$ and $(j,s)$, where $s=\ell-k\pmod m$. Lemma~\ref{lem:relabel-invariance} preserves compatibility within each family and the entries $a,b,c$. If $i=j$, interchanging both matched pairs replaces $s$ by $-s$ and preserves the tests, since the inverse adjacency sums are symmetric. If $i\ne j$, we order the two indices so that $i<j$ and use every residue $s$ modulo $m$.
		
		Consequently the reduced tests hold exactly when both full families are compatible and every matched pair satisfies the two products in~\eqref{eq:test}. Theorem~\ref{Th_mix_extension} gives the equivalence with compatibility at every repetition. Each extended path begins with its seed, so distinct seeds give $rm$ distinct paths for every $h$.
	\end{proof}
	
	The block families in Tables~\ref{tab:self_ext_8} and~\ref{tab:self_ext_10} have lengths $8$ and ten. The seed matchings in the appendix supply starting dimensions for repeated extensions. In particular, the length-$10$ blocks extend the seeds in each of the five even residue classes modulo ten. The project repository contains the construction code~\cite{PSGithub}.
	
	\begin{theorem}\label{thm:$18$}
		There is a compatible set of size $18$ in every even dimension $n\ge10$ except $n=14$. Moreover, there are compatible sets of size $15$ in dimensions $8$ and~$14$.
	\end{theorem}
	\begin{proof}
		The dimension-10 family in Table~\ref{tab:self_ext_10} supplies the blocks of length $10$. For seed dimension $n=10$, each block is matched with itself. For $n=12,16,18,24$, the matching is specified in Tables~\ref{tab:seed12}, \ref{tab:seed16}, \ref{tab:seed18}, and~\ref{tab:seed24}, respectively. The matching in dimension $14$ uses the seeds in Table~\ref{tab:seed14} and the blocks of length $8$ in Table~\ref{tab:self_ext_8}.
		
		For a pair of matched indices, let $A$ be the seed adjacency sum and $B$ the block adjacency sum. Row reduction of $[A\mid I_d]$ and $[B\mid I_t]$ first checks their ranks. The entries $a,b,c$ can then be read from the right blocks as in~\eqref{eq:abc}. For example, for the first two seeds in row $k=0$ of Table~\ref{tab:seed12}, the last seed endpoints are two and one. Their block first endpoints are $10$ and $8$, and both block last endpoints are one. The solution of $Ax=e_1$ has support $\{3,4,5,9,10,11,12\}$, so $a=x_2=0$. The solution of $By=e_8$ has support $\{1,2,5,6,7,9,10\}$, so $b=y_{10}=1$. The inverse of $B$ is alternating by Lemma~\ref{lem:offset}, so $c=(B^{-1})_{1,1}=0$.
		
		In each matching the seed sums have rank $d$ and the block sums have rank $t$. The table records the number of pairs checked and the number for which $b=1$. For each of those pairs both $a$ and $c$ are zero, as in the calculation above. The remaining pairs satisfy $ab=bc=0$ because $b=0$.
		
		\[
		\begin{array}{|c|r|r|r|}
			\hline
			n&t&\text{pairs checked}&\text{pairs with }b=1\\\hline
			10,12,16,18,24&10&153\text{ for each }n&36\text{ for each }n\\
			14&8&105&40\\
			\hline
		\end{array}
		\]
		Theorem~\ref{Th_mix_extension} gives $18$ compatible permutations in every dimension $n+10k$ for $n\in\{10,12,16,18,24\}$ and $k\ge0$. The starting dimensions represent the residues $0,2,6,8,4$ modulo ten, respectively. Together the progressions cover the even dimensions $n\ge10$ with the sole exception of $14$, where the matched seed has size $15$. The family in Table~\ref{tab:self_ext_8} gives $15$ in dimension $8$.
	\end{proof}

	\section{Compatible sets in odd dimensions}
	\label{sec:odd}
	
	We now pass from even-dimensional compatible sets to odd-dimensional compatible sets.  Let $n$ be even and $\pi=[\pi(1),\ldots,\pi(n)]\in S_n$. Consider $m=1$ and denote the extended permutations of $\pi$ with respect to $\rho=[1]\in S_m$ as $\Ext(\pi) = \{
	[\pi(1),\ldots,\pi(n),n+1], [n+1,\pi(1),\ldots,\pi(n)]
	\}$.
	Attaching one new vertex at an endpoint preserves the old adjacency sum as a principal submatrix. This gives the required rank in the next odd dimension.
	
	\begin{proposition}\label{thm:oddlift}
		Let $n$ be even and let $\pi$ and $\sigma$ be compatible permutations in $S_n$.
		Then the extended permutations $\widehat{\pi} \in \Ext(\pi)$ and 
		$\widehat{\sigma} \in \Ext(\sigma)$ are compatible.
	\end{proposition}
	\begin{proof}
		The position of the new vertex does not change the edges between the old vertices. By Lemma~\ref{lem_adjmat}, the principal submatrix of $A_{\widehat{\pi}}+A_{\widehat{\sigma}}$ indexed by $[n]$ is $A_\pi+A_\sigma$. This submatrix is nonsingular because $\pi$ and $\sigma$ are compatible and $n$ is even. The enlarged sum therefore has rank at least $n$. It is alternating of odd order $n+1$, so its rank is even and at most $n$. Its rank is exactly $n$, which gives near-bentness of $Q_{\widehat\pi}+Q_{\widehat\sigma}$ by Lemma~\ref{prop:rank}.
	\end{proof}
	
	Since the endpoint can be chosen independently for each path, the pairwise lifting result extends to the entire family.
	
	\begin{corollary}\label{cor:oddmain}
		Let $n$ be even and let $S\subseteq S_n$ be a compatible set of size $L$. For each $\pi\in S$, we choose one $\widehat{\pi}\in\Ext(\pi)$. The resulting set $\widehat S$ is compatible in $S_{n+1}$ and has size $L$. In particular, $\Mb(n+1)\ge\Mb(n)$.
	\end{corollary}
	\begin{proof}
		For every distinct pair of seed paths, Proposition~\ref{thm:oddlift} gives compatibility after either choice of attachment at each endpoint. Removing the new vertex recovers each seed, so the lifted family has size $L$. Taking an even-dimensional family of maximum size gives the inequality.
	\end{proof}
	
	The one-vertex extension transfers the even-dimensional bounds to odd dimensions at least nine. A separate construction in dimension $7$ completes the bounds for odd dimensions.
	
	\begin{corollary}\label{cor:uniform15}
		For every $n\ge7$, one has $\Mb(n)\ge15$. In dimension $7$, $\Mb(7)\ge28$. For $n\ge10$ with $n\notin\{14,15\}$, one has $\Mb(n)\ge18$.
	\end{corollary}
	\begin{proof}
		Theorem~\ref{thm:$18$} gives the bounds in even dimensions $n\ge8$, and Corollary~\ref{cor:oddmain} transfers them to the following odd dimensions. For $n=7$ let $g=(1\,2\,3\,4\,5\,6\,7)$ and take the seven translates under $g$ of each of the four paths
		\[
		[1,6,7,2,5,3,4],\quad [1,4,5,7,2,3,6],\quad
		[1,2,7,4,6,5,3],\quad [1,2,5,4,3,6,7].
		\]
		The four orbits are disjoint. As in Proposition~\ref{prop:orbit-reduction}, simultaneous relabeling reduces the rank checks to the relative exponent $h$. For each of the six pairs of distinct representatives, the seven adjacency sums at $0\le h<7$ all have rank six over $\Ftwo$. Within each orbit the three sums at $h=1,2,3$ also have rank six, and interchanging paths covers $h=4,5,6$. These $6\cdot7+4\cdot3=54$ comparisons give near-bentness of every pairwise sum by Lemma~\ref{prop:rank}.
	\end{proof}
	
	Paterson~\cite{PatersonSETA2002} bounded $\Mb(n)$ in even dimension by counting possible first rows of path adjacency matrices. In odd dimension, we count pairs of rows corresponding to the endpoints of an edge. 
	
	\begin{lemma}
		\label{lem:odd-two-row-injectivity}
		Let $n\ge 5$ be odd and let $S\subseteq S_n$ be a compatible set.  For fixed distinct vertices $a,b\in[n]$, write $N_{\Gamma_\pi}(a)$ for the set of neighbors of $a$ in $\Gamma_\pi$. Then the map
		\[
		\pi\longmapsto \bigl(N_{\Gamma_\pi}(a),N_{\Gamma_\pi}(b)\bigr)
		\]
		is injective on $S$.
	\end{lemma}
	
	\begin{proof}
		If two distinct permutations $\pi,\sigma\in S$ have the same neighbor sets at
		both $a$ and $b$, then the $a$-th and $b$-th rows of $A_\pi+A_\sigma$ are zero.
		Since the matrix is symmetric, the corresponding columns are zero as well.
		Hence $\rank(A_\pi+A_\sigma)\le n-2$.  But near-bentness of
		$Q_\pi+Q_\sigma$ in odd dimension $n$ is equivalent, by
		Lemma~\textup{\ref{prop:rank}}, to
		$\rank(A_\pi+A_\sigma)=n-1$, a contradiction.
	\end{proof}
	
	Counting incidences between paths and their edges now gives a cubic bound.
	\begin{proposition}\label{prop:oddcubic}
		For odd $n\ge5$, one has $\Mb(n)\le\binom n2(n-2)$.
	\end{proposition}
	
	\begin{proof}
		Fix an edge $e=\{a,b\}$. By Lemma~\ref{lem:odd-two-row-injectivity}, the ordered pair of neighbor sets at $a,b$ identifies each member of the compatible set that contains $e$.
		
		If exactly one of $a,b$ is a leaf, there are $2(n-2)$ possible ordered pairs of neighbor sets. If both have degree two, their additional neighbors are distinct because a path has no triangle. This gives $(n-2)(n-3)$ choices. The edge cannot join two leaves when $n>2$. Thus at most $(n-1)(n-2)$ paths in the family contain $e$. Summing this bound over all $\binom n2$ edges counts each path exactly $n-1$ times. Dividing by $n-1$ gives the result.
	\end{proof}
	
	\section{Conclusion}\label{sec:6}
	
	We used a Schur complement to reduce compatibility under repeated right extensions to two products of inverse endpoint entries for each pair of paths. By matching seeds with blocks, we constructed compatible sets of size $18$ in all even dimensions $n\ge10$ except $14$, and sets of size $15$ in dimensions $8$ and $14$. The one-vertex lift carries the bounds to the next odd dimension. A separate construction gives $28$ compatible permutations in dimension $7$. In dimension six, exhaustive enumeration gives the exact maxima of $9$ for compatibility and $8$ for compatibility under every self-extension. Together these results give at least nine compatible permutations in every dimension $n\ge6$. For odd $n\ge5$, the edge-incidence count gives $\Mb(n)\le\binom n2(n-2)$. As applications, each derived compatible set of size $L\ge2$ gives $L2^n$ spreading sequences of PAPR at most $2$ and coherence $2^{-n/2}$ in even dimension or $2^{-(n-1)/2}$ in odd dimension. In even dimension the sequences form $L$ mutually unbiased bases.
	
	\vspace{.3cm}
	
	\noindent
	\textbf{Declaration on the use of AI.}
	The authors used OpenAI  tools to assist with manuscript proofreading and the preparation of Python code for finding and checking the seeds and blocks in Section~\ref{sec:constructions}. They checked the programs and take responsibility for all the content of the paper and the accuracy of the code.

	\section*{Appendix. Seed permutations}
	\label{seeds}
	\leavevmode\par
	
	The seed $\pi_{i,k}$ is matched with the block $\omega^k(\rho_i)$ from Table~\ref{tab:self_ext_10} for $t=10$ or from Table~\ref{tab:self_ext_8} for $t=8$. In the two-column seed tables, the column and row specify $i$ and $k$, respectively.
	
	\begin{table}[b]
		\centering\small
		\caption{Seeds in dimension $n=12$ for the $18$ blocks in $S_{10}$.}\label{tab:seed12}
		\setlength{\tabcolsep}{4pt}
		\begin{tabular}{@{}ccc@{}}
			\toprule
			$k$ & $\pi_{1,k}$ & $\pi_{2,k}$\\
			\midrule
			$0$ & $[1,11,12,4,5,8,6,7,9,3,10,2]$ & $[3,8,2,11,4,12,10,6,5,9,7,1]$\\
			$1$ & $[2,11,12,5,6,9,7,8,1,4,10,3]$ & $[4,9,3,11,5,12,10,7,6,1,8,2]$\\
			$2$ & $[3,11,12,6,7,1,8,9,2,5,10,4]$ & $[5,1,4,11,6,12,10,8,7,2,9,3]$\\
			$3$ & $[4,11,12,7,8,2,9,1,3,6,10,5]$ & $[6,2,5,11,7,12,10,9,8,3,1,4]$\\
			$4$ & $[5,11,12,8,9,3,1,2,4,7,10,6]$ & $[7,3,6,11,8,12,10,1,9,4,2,5]$\\
			$5$ & $[6,11,12,9,1,4,2,3,5,8,10,7]$ & $[8,4,7,11,9,12,10,2,1,5,3,6]$\\
			$6$ & $[7,11,12,1,2,5,3,4,6,9,10,8]$ & $[9,5,8,11,1,12,10,3,2,6,4,7]$\\
			$7$ & $[8,11,12,2,3,6,4,5,7,1,10,9]$ & $[1,6,9,11,2,12,10,4,3,7,5,8]$\\
			$8$ & $[9,11,12,3,4,7,5,6,8,2,10,1]$ & $[2,7,1,11,3,12,10,5,4,8,6,9]$\\
			\bottomrule
		\end{tabular}
	\end{table}
	
	\begin{table}[b]
		\centering\footnotesize
		\caption{Seeds in dimension $n=16$ for the $18$ blocks in $S_{10}$.}\label{tab:seed16}
		\setlength{\tabcolsep}{4pt}
		\begin{tabular}{@{}ccc@{}}
			\toprule
			$k$ & $\pi_{1,k}$ & $\pi_{2,k}$\\
			\midrule
			$0$ & $[1,6,9,13,14,10,4,7,12,15,5,3,2,8,11,16]$ & $[12,5,1,6,4,11,2,13,14,9,16,8,15,3,7,10]$\\
			$1$ & $[10,13,15,1,11,7,14,4,16,12,3,2,5,6,8,9]$ & $[13,1,12,5,11,15,6,16,4,8,9,14,3,2,10,7]$\\
			$2$ & $[6,2,3,1,7,8,15,10,5,14,12,9,13,16,11,4]$ & $[7,13,12,14,8,2,6,4,15,3,11,1,16,10,9,5]$\\
			$3$ & $[2,4,7,13,14,11,5,8,10,15,6,1,3,9,12,16]$ & $[10,6,2,4,5,12,3,13,14,7,16,9,15,1,8,11]$\\
			$4$ & $[11,13,15,2,12,8,14,5,16,10,1,3,6,4,9,7]$ & $[13,2,10,6,12,15,4,16,5,9,7,14,1,3,11,8]$\\
			$5$ & $[4,3,1,2,8,9,15,11,6,14,10,7,13,16,12,5]$ & $[8,13,10,14,9,3,4,5,15,1,12,2,16,11,7,6]$\\
			$6$ & $[3,5,8,13,14,12,6,9,11,15,4,2,1,7,10,16]$ & $[11,4,3,5,6,10,1,13,14,8,16,7,15,2,9,12]$\\
			$7$ & $[12,13,15,3,10,9,14,6,16,11,2,1,4,5,7,8]$ & $[13,3,11,4,10,15,5,16,6,7,8,14,2,1,12,9]$\\
			$8$ & $[5,1,2,3,9,7,15,12,4,14,11,8,13,16,10,6]$ & $[9,13,11,14,7,1,5,6,15,2,10,3,16,12,8,4]$\\
			\bottomrule
		\end{tabular}
	\end{table}
	
	\begin{table}[hb]
		\centering\footnotesize
		\caption{Seeds in dimension $n=18$ for the $18$ blocks in $S_{10}$.}\label{tab:seed18}
		\setlength{\tabcolsep}{4pt}
		\resizebox{\linewidth}{!}{\begin{tabular}{@{}ccc@{}}
				\toprule
				$k$ & $\pi_{1,k}$ & $\pi_{2,k}$\\
				\midrule
				$0$ & $[17,3,14,16,5,13,10,6,1,8,7,11,9,18,12,15,2,4]$ & $[9,1,15,7,17,14,13,2,6,5,8,11,3,12,10,4,16,18]$\\
				$1$ & $[18,4,15,17,6,14,11,7,2,9,8,12,1,10,13,16,3,5]$ & $[1,2,16,8,18,15,14,3,7,6,9,12,4,13,11,5,17,10]$\\
				$2$ & $[10,5,16,18,7,15,12,8,3,1,9,13,2,11,14,17,4,6]$ & $[2,3,17,9,10,16,15,4,8,7,1,13,5,14,12,6,18,11]$\\
				$3$ & $[11,6,17,10,8,16,13,9,4,2,1,14,3,12,15,18,5,7]$ & $[3,4,18,1,11,17,16,5,9,8,2,14,6,15,13,7,10,12]$\\
				$4$ & $[12,7,18,11,9,17,14,1,5,3,2,15,4,13,16,10,6,8]$ & $[4,5,10,2,12,18,17,6,1,9,3,15,7,16,14,8,11,13]$\\
				$5$ & $[13,8,10,12,1,18,15,2,6,4,3,16,5,14,17,11,7,9]$ & $[5,6,11,3,13,10,18,7,2,1,4,16,8,17,15,9,12,14]$\\
				$6$ & $[14,9,11,13,2,10,16,3,7,5,4,17,6,15,18,12,8,1]$ & $[6,7,12,4,14,11,10,8,3,2,5,17,9,18,16,1,13,15]$\\
				$7$ & $[15,1,12,14,3,11,17,4,8,6,5,18,7,16,10,13,9,2]$ & $[7,8,13,5,15,12,11,9,4,3,6,18,1,10,17,2,14,16]$\\
				$8$ & $[16,2,13,15,4,12,18,5,9,7,6,10,8,17,11,14,1,3]$ & $[8,9,14,6,16,13,12,1,5,4,7,10,2,11,18,3,15,17]$\\
				\bottomrule
		\end{tabular}}
	\end{table}

	\begin{table}[ht]
		\centering\small
		\caption{Seeds in dimension $n=24$ for the $18$ blocks in $S_{10}$.}\label{tab:seed24}
		\setlength{\tabcolsep}{4pt}
		\begin{tabular}{@{}ccl@{}}
			\toprule
			$k$ & block & matched seed $\pi_{i,k}$\\
			\midrule
			$0$ & $\omega^{0}(\rho_1)$ & $[2,24,14,17,11,10,18,7,20,19,6,15,22,8,16,12,9,1,23,21,13,5,4,3]$\\
			& $\omega^{0}(\rho_2)$ & $[10,8,4,3,9,2,7,14,24,15,16,20,21,1,5,23,22,6,12,13,17,19,11,18]$\\
			\addlinespace[2pt]
			$1$ & $\omega^{1}(\rho_1)$ & $[3,22,15,18,12,11,10,8,21,20,7,16,23,9,17,13,1,2,24,19,14,6,5,4]$\\
			& $\omega^{1}(\rho_2)$ & $[11,9,5,4,1,3,8,15,22,16,17,21,19,2,6,24,23,7,13,14,18,20,12,10]$\\
			\addlinespace[2pt]
			$2$ & $\omega^{2}(\rho_1)$ & $[4,23,16,10,13,12,11,9,19,21,8,17,24,1,18,14,2,3,22,20,15,7,6,5]$\\
			& $\omega^{2}(\rho_2)$ & $[12,1,6,5,2,4,9,16,23,17,18,19,20,3,7,22,24,8,14,15,10,21,13,11]$\\
			\addlinespace[2pt]
			$3$ & $\omega^{3}(\rho_1)$ & $[5,24,17,11,14,13,12,1,20,19,9,18,22,2,10,15,3,4,23,21,16,8,7,6]$\\
			& $\omega^{3}(\rho_2)$ & $[13,2,7,6,3,5,1,17,24,18,10,20,21,4,8,23,22,9,15,16,11,19,14,12]$\\
			\addlinespace[2pt]
			$4$ & $\omega^{4}(\rho_1)$ & $[6,22,18,12,15,14,13,2,21,20,1,10,23,3,11,16,4,5,24,19,17,9,8,7]$\\
			& $\omega^{4}(\rho_2)$ & $[14,3,8,7,4,6,2,18,22,10,11,21,19,5,9,24,23,1,16,17,12,20,15,13]$\\
			\addlinespace[2pt]
			$5$ & $\omega^{5}(\rho_1)$ & $[7,23,10,13,16,15,14,3,19,21,2,11,24,4,12,17,5,6,22,20,18,1,9,8]$\\
			& $\omega^{5}(\rho_2)$ & $[15,4,9,8,5,7,3,10,23,11,12,19,20,6,1,22,24,2,17,18,13,21,16,14]$\\
			\addlinespace[2pt]
			$6$ & $\omega^{6}(\rho_1)$ & $[8,24,11,14,17,16,15,4,20,19,3,12,22,5,13,18,6,7,23,21,10,2,1,9]$\\
			& $\omega^{6}(\rho_2)$ & $[16,5,1,9,6,8,4,11,24,12,13,20,21,7,2,23,22,3,18,10,14,19,17,15]$\\
			\addlinespace[2pt]
			$7$ & $\omega^{7}(\rho_1)$ & $[9,22,12,15,18,17,16,5,21,20,4,13,23,6,14,10,7,8,24,19,11,3,2,1]$\\
			& $\omega^{7}(\rho_2)$ & $[17,6,2,1,7,9,5,12,22,13,14,21,19,8,3,24,23,4,10,11,15,20,18,16]$\\
			\addlinespace[2pt]
			$8$ & $\omega^{8}(\rho_1)$ & $[1,23,13,16,10,18,17,6,19,21,5,14,24,7,15,11,8,9,22,20,12,4,3,2]$\\
			& $\omega^{8}(\rho_2)$ & $[18,7,3,2,8,1,6,13,23,14,15,19,20,9,4,22,24,5,11,12,16,21,10,17]$\\
			\bottomrule
		\end{tabular}
	\end{table}

	\begin{table}[ht]
		\centering
		\caption{Seeds in dimension $n=14$ for the $15$ blocks in $S_8$}\label{tab:seed14}
		\setlength{\tabcolsep}{4pt}
		\begin{tabular}{@{}ccl@{}}
			\toprule
			$k$ & block & matched seed $\pi_{i,k}$\\
			\midrule
			$0$ & $\omega^{0}(\rho_1)$ & $[5,10,8,14,6,4,12,9,7,2,3,11,13,1]$\\
			& $\omega^{0}(\rho_2)$ & $[2,8,6,4,13,1,11,3,9,12,10,14,5,7]$\\
			& $\omega^{0}(\rho_3)$ & $[13,4,10,1,9,7,12,3,5,6,11,14,2,8]$\\
			\addlinespace[2pt]
			$1$ & $\omega^{1}(\rho_1)$ & $[1,6,9,14,7,5,12,10,8,3,4,11,13,2]$\\
			& $\omega^{1}(\rho_2)$ & $[3,9,7,5,13,2,11,4,10,12,6,14,1,8]$\\
			& $\omega^{1}(\rho_3)$ & $[13,5,6,2,10,8,12,4,1,7,11,14,3,9]$\\
			\addlinespace[2pt]
			$2$ & $\omega^{2}(\rho_1)$ & $[2,7,10,14,8,1,12,6,9,4,5,11,13,3]$\\
			& $\omega^{2}(\rho_2)$ & $[4,10,8,1,13,3,11,5,6,12,7,14,2,9]$\\
			& $\omega^{2}(\rho_3)$ & $[13,1,7,3,6,9,12,5,2,8,11,14,4,10]$\\
			\addlinespace[2pt]
			$3$ & $\omega^{3}(\rho_1)$ & $[3,8,6,14,9,2,12,7,10,5,1,11,13,4]$\\
			& $\omega^{3}(\rho_2)$ & $[5,6,9,2,13,4,11,1,7,12,8,14,3,10]$\\
			& $\omega^{3}(\rho_3)$ & $[13,2,8,4,7,10,12,1,3,9,11,14,5,6]$\\
			\addlinespace[2pt]
			$4$ & $\omega^{4}(\rho_1)$ & $[4,9,7,14,10,3,12,8,6,1,2,11,13,5]$\\
			& $\omega^{4}(\rho_2)$ & $[1,7,10,3,13,5,11,2,8,12,9,14,4,6]$\\
			& $\omega^{4}(\rho_3)$ & $[13,3,9,5,8,6,12,2,4,10,11,14,1,7]$\\
			\bottomrule
		\end{tabular}
	\end{table}

\end{document}